\documentclass[aps,prc,twocolumn,groupedaddress,showpacs]{revtex4-1}\usepackage{bm,physics}
\usepackage[T1]{fontenc}
\usepackage{textcomp}
\usepackage[dvips,final]{graphicx}
\usepackage{ulem, color}
\usepackage{hyperref}
\bmdefine{\ba}{a}
\bmdefine{\bb}{b}
\bmdefine{\bx}{x}
\bmdefine{\by}{y}
\bmdefine{\bz}{z}
\bmdefine{\bn}{n}
\bmdefine{\bp}{p}
\newcommand{\BM}{\begin{pmatrix}}
\newcommand{\EM}{\end{pmatrix}}

\newif\ifHIDEHIGHLIGNT
\HIDEHIGHLIGNTtrue %%%%% highlight the alterations
\ifHIDEHIGHLIGNT
\usepackage{ulem,color}
\newcounter{nnn}

\else

\fi

\begin{document}
\title {%Neutrinoless double $\beta$ decay of 
$\alpha$ + $^{92}$Zr cluster structure in $^{96}$Mo}

\author{
S. Ohkubo$^{1}$ and Y. Hirabayashi$^2$ 
}
\affiliation{$^1$ Research Center for Nuclear Physics, Osaka University, 
Ibaraki, Osaka 567-0047, Japan }
\affiliation{$^2$Information Initiative Center,
Hokkaido University, Sapporo 060-0811, Japan}

\date{\today}
\begin{abstract}
\par
In the evaluation of the half-life of the neutrinoless double-$\beta$ decay ($0\nu\beta\beta$) of a doubly closed-subshell nucleus $^{96}$Zr, the structure of the nucleus $^{96}$Mo is essentially important. The $\alpha$-clustering aspects of $^{96}$Mo
are investigated for the first time. By studying the nuclear rainbows in $\alpha$ scattering from $^{92}$Zr at high energies and the characteristic structure of the excitation functions at the extreme backward angle at the low-energy region, the interaction potential between the $\alpha$ particle and the $^{92}$Zr nucleus is determined well in the double folding model. The validity of the double folding model was reinforced by studying $\alpha$ scattering from neighboring nuclei $^{90}$Zr, $^{91}$Zr, and $^{94}$Zr. The double-folding-model calculations reproduced  well all the observed angular distributions over a wide range of incident energies and the characteristic excitation functions. By using the obtained potential the $\alpha$ +$^{92}$Zr cluster structure of $^{96}$Mo is investigated in the spirit of a unified description of scattering and structure. The existence of the second-higher nodal band states with the $\alpha$+ $^{92}$Zr cluster structure, in which two more nodes are excited in the relative motion compared with the ground band, is demonstrated. The calculation reproduces well the ground-band states of $^{96}$Mo in agreement with experiment. The experimental $B(E2)$ value of the transition in the ground band is also reproduced well. The effect of $\alpha$ clustering in $^{96}$Mo on the the half-life of the $0\nu\beta\beta$ double-$\beta$ decay of $^{96}$Zr is discussed.
\end{abstract}

\maketitle
\par
\section{INTRODUCTION}

\par
The observation of neutrinoless double-$\beta$ decay, $0\nu\beta\beta$, which violates lepton number conservation, is expected to serve to shed light on the fundamental questions beyond the standard model, such as determining the nature of neutrino, Dirac, or Majorana particles. Since supersymmetric particles have not been observed in Large  Hadron Collider experiments, much more attention than ever has been paid to study of $0\nu\beta\beta$ \cite{Ejiri2005,Avignone2008,Vergados2012}. 
The inverse half-life of $0\nu\beta\beta$ 
is given by 
$ [T^{0\nu}_{1/2}]^{-1}=G_{0\nu} \left|{<m_{\beta\beta}>/}{m_e}\right|^2 |M^{0\nu}|^2$,
where $<m_{\beta\beta}>$ is the effective Majorana neutrino mass,
${m_e}$ is the electron mass, and $G_{0\nu}\sim 10^{-14}$ yr$^{-1}$ is a phase-space factor. For the evaluation of the 
nuclear matrix element (NME) of the transition $M^{0\nu}$ \cite{Suhonen1998,Faessler1998,Suhonen2012,Engel2017,%
Jokiniemi2018,Dolinski2019}, 
it is essential to know the ground-state wave functions of the initial- and final- state nuclei.
\par
Up to now theoretical $0\nu\beta\beta$ decay study has been done based on the mean-field theory such as the shell-model \cite{Iwata2016,Coraggio2020}, {\it ab initio} calculations \cite{Yao2020,Belley2021}, quasiparticle random phase approximation (QRPA) \cite{Simkovic2013,Hyvarinen2015,Jokiniemi2021}, the projected Hartree-Fock Bogoliubov model (PHFB) \cite{Chaturvedi2008,Rath2010,Nautiyal2022}, the generator coordinate method (GCM) \cite{Rodriguez2010,Hinohara2014,Song2017,Jiao2019}, the energy density functional (EDF) \cite{Vaquero2013,Yao2015} and the interacting boson model (IBM) \cite{Barea2009,Barea2013}.
No attention has been paid to the $\alpha$ cluster structure viewpoint until the study of $^{48}$Ca decay to $^{48}$Ti \cite{Ohkubo2021}. This is probably because it has been believed intuitively that strong spin-orbit force would break $\alpha$ clustering and partly because experimental data of $\alpha$-transfer reactions such as ($^6$Li,d), (d,$^6$Li), and $(p,p\alpha)$ are scarce. 

\par
$\alpha$ cluster structure has been established in the light mass region \cite{Suppl1972,Suppl1980} and medium-weight mass region around $^{44}$Ti\cite{Michel1998,Yamaya1998,Sakuda1998,Ohkubo1999} and recently extended to the $^{52}$Ti region \cite{Fukada2009,Bailey2019,Ohkubo2020,Ohkubo2021,Bailey2021}. 
In a previous paper \cite{Ohkubo2021}, paying attention to the $0\nu\beta\beta$ decay of 
$^{48}$Ca to $^{48}$Ti, one of the present authors (S.O.) has shown that the ground $0^+$ state of $^{48}$Ti has $\alpha$-clustering aspects, which significantly quenches the half-life than the conventional shell model calculations in which excitations to the higher several major shells are not considered.

\par
In the $0\nu\beta\beta$ of the parent nucleus $^{96}$Zr \cite{Fukuda2020},
the structure of the ground state of the daughter nucleus $^{96}$Mo, whose $\alpha$ threshold energy 2.76 MeV is small, is crucial in evaluating the NME of $0\nu\beta\beta$ decay transitions.
The persistency of $\alpha$ clustering in the heavier mass region around $A=90$ has been explored for the typical nucleus $^{94}$Mo with two protons and two neutrons outside the closed shell core $^{90}$Zr in Refs. \cite{Ohkubo1995,Buck1995,Michel2000}. Later $\alpha$ cluster model study \cite{Ohkubo2009,Souza2015,Ni2011} also supports $\alpha$ clustering  in the $^{94}$Mo region. Recent observations of $\alpha$ particles in the pick-up reactions $(p,p\alpha)$ in the Sn isotopes \cite{Tanaka2021} seem to reinforce the importance of $\alpha$ clustering in the heavy mass region.

\par
The ground state of $^{96}$Zr is spherical being a doubly closed-subshell nucleus and is analog to the doubly closed shell $^{16}$O in light nuclei \cite{Molnar1986}. The first excited $0^+$ state is considered to be a four-particle four-hole excited-state analog to the mysterious $0^+$ state at 6.05 MeV in $^{16}$O. Recent large-scale shell-model calculations for the Zr isotopes \cite{Togashi2016} have confirmed that the ground state of $^{96}$Zr is spherical and that shape transition to deformed occurs at $^{100}$Zr as the number of the excess neutrons increases.
As for the structure of $^{96}$Mo, studies including $2\nu\beta\beta$ decay using QRPA \cite{Stoica1995}, phase transition from spherical $^{92}$Mo to deformed toward $^{104}$Mo \cite{Lesher2007}, octupole collective motion \cite{Gregor2017} and shell-model structure \cite{Coraggio2018} have been reported. $0\nu\beta\beta$ of $^{96}$Zr has been investigated  using many models, which includes the QRPA, PHFB, EDF, IBM, and GCM \cite{Nautiyal2022}. However, no study of $0\nu\beta\beta$ of $^{96}$Zr from the viewpoint of $\alpha$ cluster of $^{96}$Mo has been challenged. 

\par
The purpose of this paper is to show that $\alpha$ clustering persists in the ground state of $^{96}$Mo by studying bound states and scattering for the $\alpha$+$^{92}$Zr system in a unified way and that the half-life of $0\nu\beta\beta$ of $^{96}$Zr is quenched significantly. For this, by using a double folding model the interaction potential between $\alpha$ particle and $^{92}$Zr is determined by analyzing angular distributions of nuclear rainbows in $\alpha$+$^{92}$Zr scattering at high energies, backward angle anomaly (BAA) or anomalous large angle scattering (ALAS) at lower energies. The potential reproduces well the excitation functions with a characteristic dip at the extreme backward angles near 180$^\circ$ in the lower-energy region systematically not only for $\alpha$+$^{92}$Zr scattering but also for $\alpha$+$^{90,91,94}$Zr scattering. The existence of the second-higher nodal band states with the $\alpha$+$^{92}$Zr cluster structure, which is responsible for the emergence of the characteristic dip in the back-angle excitation function, is shown for the first time. The ground band of $^{96}$Mo is understood well in the $\alpha$-cluster model study using the obtained double folding potential. $\alpha$ clustering of $^{96}$Mo gives significant effect to quench the $0\nu\beta\beta$ decay of $^{96}$Zr.

\par
The paper is organized as follows. In Sec. II the double folding model is presented. Section III is devoted to the analysis of $\alpha$+$^{92}$Zr scattering over a wide range of incident energies by using a double folding model. To confirm the validity of the obtained interaction potential for $\alpha$+$^{92}$Zr, $\alpha$ scattering from neighboring nuclei $^{90,91,94}$Zr is also investigated. 
In Sec. IV the origin of the characteristic dip in the back-angle excitation function in $\alpha$+$^{92}$Zr scattering is investigated from the viewpoint of 
persistent existence of the 
$\alpha$ cluster structure at the highly excited energies in $^{96}$Mo.
In Sec. V, $\alpha$+$^{92}$Zr clustering of $^{96}$Mo is studied and discussions of $\alpha$ clustering on the $0\nu\beta\beta$ decay of $^{96}$Zr is given. A summary is given in Sec. VI. 

\section{ DOUBLE FOLDING MODEL}
\par
We study $\alpha$ scattering from $^{92}$Zr and neighboring nuclei $^{90,91,94}$Zr  with a double folding 
model using a density-dependent nucleon-nucleon force.
The double folding potential is calculated as follows:
\begin{eqnarray}
\lefteqn{V({\bf r}) =
\int \rho_{00}^{\rm (^{4}He)} ({\bf r}_{1})\;
\rho_{00}^{\rm (^{}Zr)} ({\bf r}_{2})} \nonumber\\
&& \times v_{\it NN} (E,\rho,{\bf r}_{1} + {\bf r} - {\bf r}_{2})\;
{\it d}{\bf r}_{1} {\it d}{\bf r}_{2} ,
\label{folding}
\end{eqnarray}
\noindent where $\rho_{00}^{\rm (^{4}He)} ({\bf r_1})$ and $\rho_{00}^{\rm (^{}Zr)} ({\bf r_2})$ represent the 
nucleon density of the ground states of $^{4}$He and $^{}$Zr, respectively, which 
are obtained by the convolution of the proton size from the charge density distribution taken from Ref.\cite{DeVries1987}.
For the effective interaction $v_{\rm NN}$ we use 
the density($\rho$)-dependent M3Y interaction  \cite{Kobos1982}.
In the calculations we introduce the normalization factor $N_R$ for 
the real double folding potential \cite{Satchler1979,Brandan1997}. 
The Coulomb folding potential is calculated similarly by the folding prescription in Eq.~(\ref{folding}). 
An imaginary potential with a Woods-Saxon volume-type form factor (nondeformed) is introduced phenomenologically to take into account the effect
of absorption due to other channels.

\section{ANALYSIS OF ALPHA SCATTERING FROM $^{92}$Z\lowercase{r} and $^{90,91,94}$Z\lowercase{r} }
\par
In exploring the $\alpha$ cluster structure in the medium-weight mass region where the level density is high, a unified description of $\alpha$ scattering including rainbow scattering, prerainbows and BAA (ALAS), and the $\alpha$ cluster structure in the bound and quasibound energy region has been very powerful \cite{Michel1986,Michel1988,Ohkubo1999,Ohkubo2020,Ohkubo2021}. 

The angular distributions in $\alpha$ scattering from $^{92}$Zr have been measured systematically at $E_\alpha$=40, 65, 90 and 120 MeV in Ref. \cite{Put1977} and 35.4 MeV 
in Ref. \cite{Singh1986}. The interaction potential can be uniquely determined from the analysis of the angular distributions in the rainbow energy region, which show the Airy minimum in the lit side of the nuclear rainbow followed by the falloff of the cross sections corresponding to the darkside of the nuclear rainbow.

We started to analyze the angular distribution at the highest energy $E_\alpha$= 120 MeV to fit to the experimental angular distribution by introducing $N_R$=1.26 and a phenomenological imaginary potential with a strength parameter $W$=18.5 MeV, a radius parameter $R_W=$7.1 fm and a diffuseness parameter $a_W=$0.6 fm. Then by keeping the fixed $N_R$ =1.26  for 90 and 65 MeV and with a slightly reduced value $N_R$ =1.22 for 40 and 35.4 MeV,  all the angular distributions are easily reproduced by the calculations with a small adjustment to reduce the strength and/or diffuseness parameters of the imaginary potential with decreasing incident energies. The calculated angular distributions are in good agreement with the experimental data as displayed in Fig.~1. 
In Table I the values of the volume integral per nucleon pair $J_V$  and the rms radius $\sqrt{ <r_V^2>}$ of the double folding potential, and the parameters of the imaginary potential  together with the volume integral per nucleon pair $J_W$ and the rms radius $\sqrt{ <r_W^2>}$ are listed. The energy dependence of the volume integrals $J_V$ is reasonable, which is  consistent with the previous calculations for $\alpha$+$^{92}$Zr scattering in Refs. \cite{Ohkubo1995,Put1977}.

% Fig. 1 exp fit Farside 35.4 MeV-120 MeV alpha+92Zr scattering .
\begin{figure}[t!]
\includegraphics[width=8.6cm]{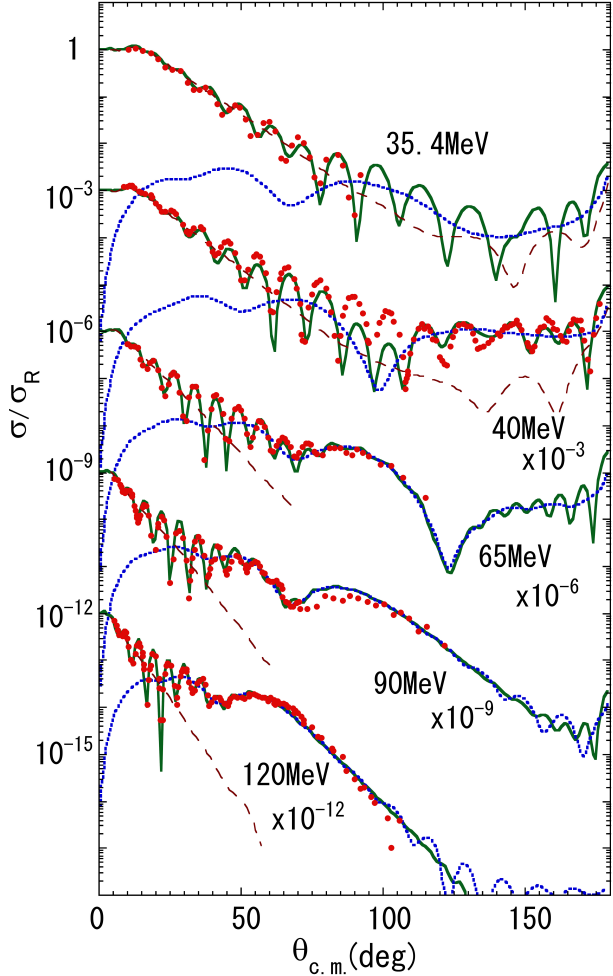}% Here is 
\protect\caption{(Color online) The angular distributions in $\alpha$+$^{92}$Zr scattering at $E_\alpha$=35.4, 40, 65, 90 and 120 MeV calculated with the optical potential model with the double folding potential (solid lines) are compared with the experimental data (filled circles) \cite{Put1977,Singh1986}. The calculated farside (dotted lines) and nearside (dashed lines) contributions are also indicated.
}
\end{figure}

\par
In order to see the contributions of the refractive farside scattering, the calculated angular distributions are decomposed into the farside and nearside components
 \cite{Fuller1975}. In Fig.~1, 
we see that the falloff of the cross sections in the angular distributions in the intermediate angular region above $E_\alpha$=65 MeV, which is peculiar to nuclear rainbow scattering, 
are all due to  farside scattering. A clear first-order Airy minimum $A1$ of the nuclear rainbow is seen at $\theta=50^\circ$ at $E_\alpha$=120 MeV, which shifts backward as the incident energy decreases, at around $\theta=70^\circ$ for $E_\alpha$=90 MeV and at $\theta=125^\circ$ for $E_\alpha$=65 MeV.
At $E_\alpha$=40 MeV no Airy minimum is observed. The appearance of the oscillations in the backward angular distributions shows that the nearside contributions are involved since the oscillations are the consequence of interference of the two amplitudes of farside and nearside scattering. 
This backward rise of the cross sections with the oscillations at $E_\alpha$=40 MeV
is the indication of BAA under incomplete absorption, which is typically observed and explained in $\alpha$+$^{16}$O \cite{Ohkubo1977,Michel1983} and $\alpha$+$^{40}$Ca scattering \cite{Delbar1978} in the energy region $E_\alpha$=20-30 MeV.

\par
In the energy region below $E_\alpha$=40 MeV the concept of farside and nearside scattering is no more powerful in understanding the characteristic features of the angular distributions. It is useful to understand the characteristics of the angular distributions of BAA in terms of the concept of internal waves and barrier waves \cite{Brink1985}.
The scattering amplitude $f(\theta)$ can be decomposed into $f^I(\theta)$, which is due to the internal waves penetrating the barrier deep into the internal region of the potential and $f^B(\theta)$, which is due to the barrier waves reflected at the barrier of the potential in the surface region, 
$f(\theta) = f^I(\theta)+ f^B(\theta)$.
In the case of incomplete absorption the internal waves, $f^I(\theta)$, carry the information of the internal region of the potential. Unfortunately at the lower energies below 30 MeV no angular distributions have been measured for $\alpha$+$^{92}$Zr scattering 
where the effect of the internal waves is clearly seen \cite{Ohkubo1977,Michel1983,Delbar1978,Brink1985}. 

% Table I potential parameters 
\begin{table}[b!]
\begin{center}
\protect\caption{  The normalization factor $N_R$, volume integral per nucleon pair $J_V$, rms radius $\sqrt{ <r_V^2>}$ of the double folding potentials, and the strength $W$, radius $R_W$, diffuseness $a_W$, volume integral per nucleon pair $J_W$ and  rms radius $\sqrt{ <r_W^2>}$ of the imaginary potentials 
used in $\alpha$+$^{92}$Zr scattering in Fig.~1. Energies are in MeV, volume integrals in MeVfm$^3$ and radii in fm.
}
\begin{tabular}{ccccccccc}
\hline
\hline
$E_\alpha$ & $N_R$ & $J_V$ & $\sqrt{ <r_V^2>}$ & $W$ & $R_W$ & $a_W$ & $J_W$ & $\sqrt{ <r_W^2>}$ \\
\hline
35.4 & 1.22& 318.9 & 5.00 & 14.0 &7.1 &0.55& 60.4 &5.87\\
40 & 1.22& 315.7 & 5.00 & 14.0 &7.1 &0.50 &59.8&5.81\\
65 & 1.26& 313.4 & 5.01 & 17.0 &7.1 &0.60 &74.1&5.94\\
90 & 1.26& 298.2 & 5.01 & 18.5 &7.1 &0.60 &80.7&5.94\\
120 & 1.26& 280.5 & 5.02 & 18.5 &7.1 &0.60 &80.7&5.94\\
\hline
\end{tabular}
\end{center}
\end{table}

%%%%Fig.2 2022.7.29 added 90Zr 91Zr angular distributions E=23-25 MeV
\begin{figure}[t]
\begin{center}
\includegraphics[width=8.6cm]{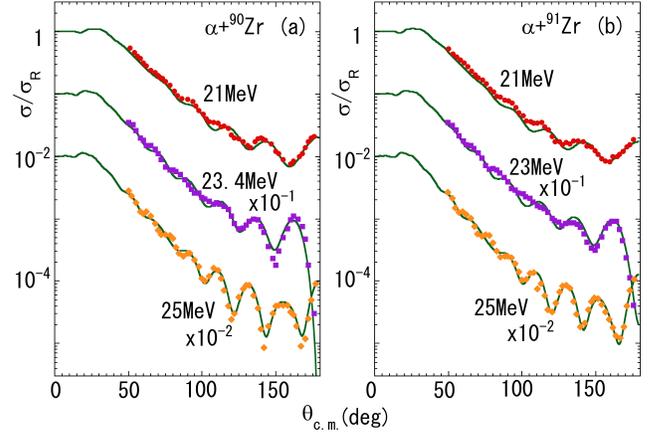}%\hspace{0.3cm} % half size
\caption{\label{fig.2} 
(Color online) The calculated angular distributions (solid lines) in  (a) $\alpha$+$^{90}$Zr scattering at $E_\alpha$= 21, 23.4 and 25 MeV and (b) $\alpha$+$^{91}$Zr 
scattering at $E_\alpha$= 21, 23, and 25 MeV are compared with the experimental data (filled circles) \cite{Wit1975}.}
\label{fig2}
\end{center}
\end{figure}
\par
However, we note that the angular distributions in $\alpha$ scattering from neighboring nuclei $^{90}$Zr and $^{91}$Zr have been measured up to the backward angles at the lower energies $E_\alpha$=23-25 MeV. In Fig.~2 the angular distributions show a BAA rising toward the extreme backward angles at 21 and 25 MeV. Note that the angular distributions for both $^{90}$Zr and $^{91}$Zr  decrease sharply toward 180$^\circ$ at $E_\alpha$=23 MeV in the BAA energy region, which is not seen in the typical $\alpha$+$^{16}$O \cite{Ohkubo1977,Michel1983} and $\alpha$+$^{40}$Ca scattering \cite{Delbar1978}. 
This characteristic decrease is intriguing because angular distributions at other energies generally increase toward $\theta=180^\circ$, see Fig.~1, as expected from the behavior of the Legendre polynomials whose moduli increases toward
$\theta=180^\circ$ at the extreme back angles.
In Fig.~2 the angular distributions in $\alpha$+ $^{90}$Zr and $\alpha$+$^{91}$Zr scattering calculated using the double folding potential derived from Eq.~(1) are compared with the experimental data \cite{Wit1975}. 
The potential parameters used are listed in Table II. The calculations reproduce the experimental angular distributions well. Note that  the particular behavior at 23 MeV that  decreases sharply toward 180$^\circ$ is reproduced excellently. 
This shows that the calculated double folding potentials for $\alpha$+ $^{90}$Zr and $\alpha$+$^{91}$Zr work very well in this low-energy region, which reinforces the validity of the double folding potential in the $E_\alpha$=23- MeV to $E_\alpha$=25- MeV region.

\par
In Fig.~3 the excitation functions at the extreme backward angle $\theta$=176.2$^\circ$ ($\theta_{Lab}$=176$^\circ$) in $\alpha$ scattering from $^{90}$Zr, $^{91}$Zr, $^{92}$Zr and $^{94}$Zr calculated using the potentials at $E_\alpha$= 23 MeV in Table II are displayed in comparison with the experimental data.
All the calculated excitation functions show a dip and its position shifts to lower energy from $^{90}$Zr to $^{94}$Zr. 
The position of the dips in the calculated excitation functions for $\alpha$+$^{90}$Zr, $\alpha$+$^{91}$Zr and $\alpha$+$^{94}$Zr agrees with the experimental data excellently. 
The energy of the observed dips for $^{90}$Zr, $^{91}$Zr and $^{94}$Zr deceases linearly with  the mass number $A$ of the target nucleus, $E_\alpha$=54.5 - 0.346$A$, which predicts a dip at $E_\alpha$=22.7 MeV for $^{92}$Zr.
As seen in Fig.~3, the double folding model calculation locates a dip at $E_\alpha$= 22.7 MeV for $\alpha$+$^{92}$Zr, which is in good agreement with the above-predicted energy, 22.7 MeV.

% Fig.3 alpha+ 92Zr 90Zr 91Zr 94Zr excitation function 
\begin{figure}[b!]
\includegraphics[width=7.0cm]{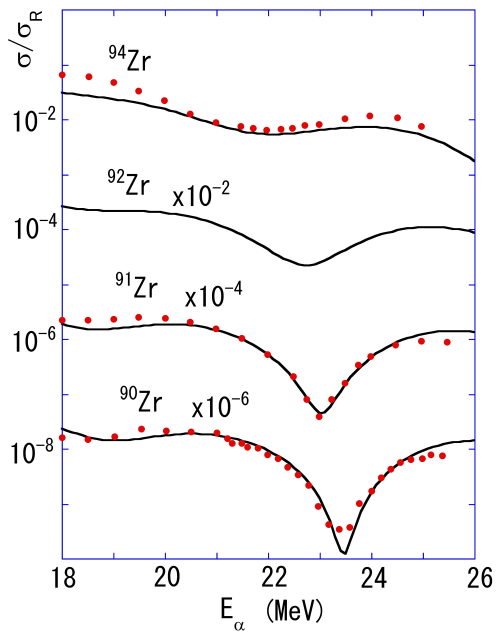}% Here is how to import EPS art
\protect\caption {(Color online) The calculated excitation functions in $\alpha$ scattering from $^{90}$Zr, $^{91}$Zr, $^{92}$Zr, and $^{94}$Zr at the extreme backward angle $\theta$=176.2$^\circ$ ($\theta_{Lab}$=176$^\circ$)
(solid lines) are compared with the experimental data (filled circles)  \cite{Wit1975}. }
\label{fig3}
\end{figure}

%% Table II potential parameters E_alpha=21, 23, 25 MeV 90Zr 91Zr 94Zr
\begin{table}[t!]
\begin{center}
\protect\caption{  The volume integral per nucleon pair $J_V$, rms radius $\sqrt{ <r_V^2>}$ of the double folding potentials, and the strength $W$, radius $R_W$, diffuseness $a_W$, volume integral per nucleon pair $J_W$, and  rms radius $\sqrt{ <r_W^2>}$ of the imaginary potentials 
used in $\alpha$+$^{90,91,92,94}$Zr scattering in Fig.~2 and Fig.~3. Energies are in MeV, volume integrals in MeVfm$^3$, and radii in fm. $N_R$=1.22 is used for all target nuclei and incident energies. 
}
\begin{tabular}{clccccccc}
\hline
\hline
& $E_\alpha$ &  $J_V$ & $\sqrt{ <r_V^2>}$ & $W$ & $R_W$ & $a_W$ & $J_W$ & $\sqrt{ <r_W^2>}$ \\
\hline
$^{90}$Zr & 21 &  331.5 &4.96 & 10.0 &7.55 &0.40 & 51.5 &6.03\\
& 23.4 & 329.5 &4.96 & 10.0 &7.55 &0.43 & 51.7&6.06\\
& 25 &  327.5 &4.96 & 10.0 &7.55 &0.48 & 52.1&6.11\\ 
$^{91}$Zr & 21 &  333.1 &5.00 & 10.6 &7.60 &0.37 & 54.8&6.05\\
& 23 &  331.7 &5.00 & 10.2 &7.60 &0.41 & 53.0&6.08\\
&25 &  330.3 &5.01 & 10.2 &7.60 &0.45 & 53.3&6.12\\ 
$^{92}$Zr & 23 &  329.6 &4.99 & 10.7 &7.63 &0.43 &55.8&6.12\\
$^{94}$Zr & 23 &  330.0 &5.02 & 11.8 &7.70 &0.48 & 62.3&6.23\\
\hline
\end{tabular}
\end{center}
\end{table}

\par
The mechanism explaining why the dip emerges in the excitation function at the extreme backward angle near $\theta$=180$^\circ$, namely why the angular distribution decreases sharply toward $\theta$=180$^\circ$ at a particular energy, has been investigated in detail for the typical $\alpha$+$^{90}$Zr system by one of the present authors (S.O.) and his collaborators, see Ref. \cite{Michel2000}. The mechanism is understood as follows.
The dip appears at the energy where the scattering amplitude $f(\theta)$ becomes vanishingly small. 
When $f^I(\theta)$$\approx$$- f^B(\theta)$, the cancellation of the two amplitude occurs, i.e., in the case when 
$|f^I(\theta)|$ $\approx$ $| f^B(\theta)|$ and
arg($|f^I(\theta))$- arg($ f^B(\theta))$ $\approx$ $k\pi$ where $k$ is an odd integer. 
At near $\theta$=$180^\circ$ this condition is satisfied at the energy $E_\alpha$=22-24 MeV under moderate absorption not only for 
$\alpha$+$^{90}$Zr but also for $\alpha$+$^{91}$Zr, $\alpha$+$^{92}$Zr and $\alpha$+$^{94}$Zr since both the real potential and the imaginary potential change little from that of
$\alpha$+$^{90}$Zr as seen in Table II. The good agreement of the calculated excitation functions, especially the energy position and width of the dip for $\alpha$+$^{91}$Zr, and $\alpha$+$^{94}$Zr with the experimental data, is the natural consequence that their potentials resemble that for $\alpha$+$^{90}$Zr. Although no experimental data are available for $\alpha$+$^{92}$Zr, the emergence of the dip at the predicted energy in the excitation function 
would be confirmed in the future experiment. 
Since the internal waves, which are responsible for the emergence of the dip, are sensitive to the internal region of the real potential, the present good agreement in Fig.~3 shows that the obtained double folding potential is reliable sufficiently in this low-energy region above the Coulomb barrier.

\section{MECHANISM OF THE CHARACTERISTIC DIP IN THE BACK-ANGLE EXCITATION FUNCTION IN $\alpha$+$^{92}$Z\lowercase{r} SCATTERING}

  In this section, paying attention to the highly lying excited $\alpha$ cluster structure in $^{96} $Mo, we investigate  how the anomalous dip in the back-angle excitation in $\alpha$+$^{92}$Z\lowercase{r} scattering  in Fig.~3  is created.
 
 \par
\begin{figure}[t!]
\includegraphics[width=7.0cm]{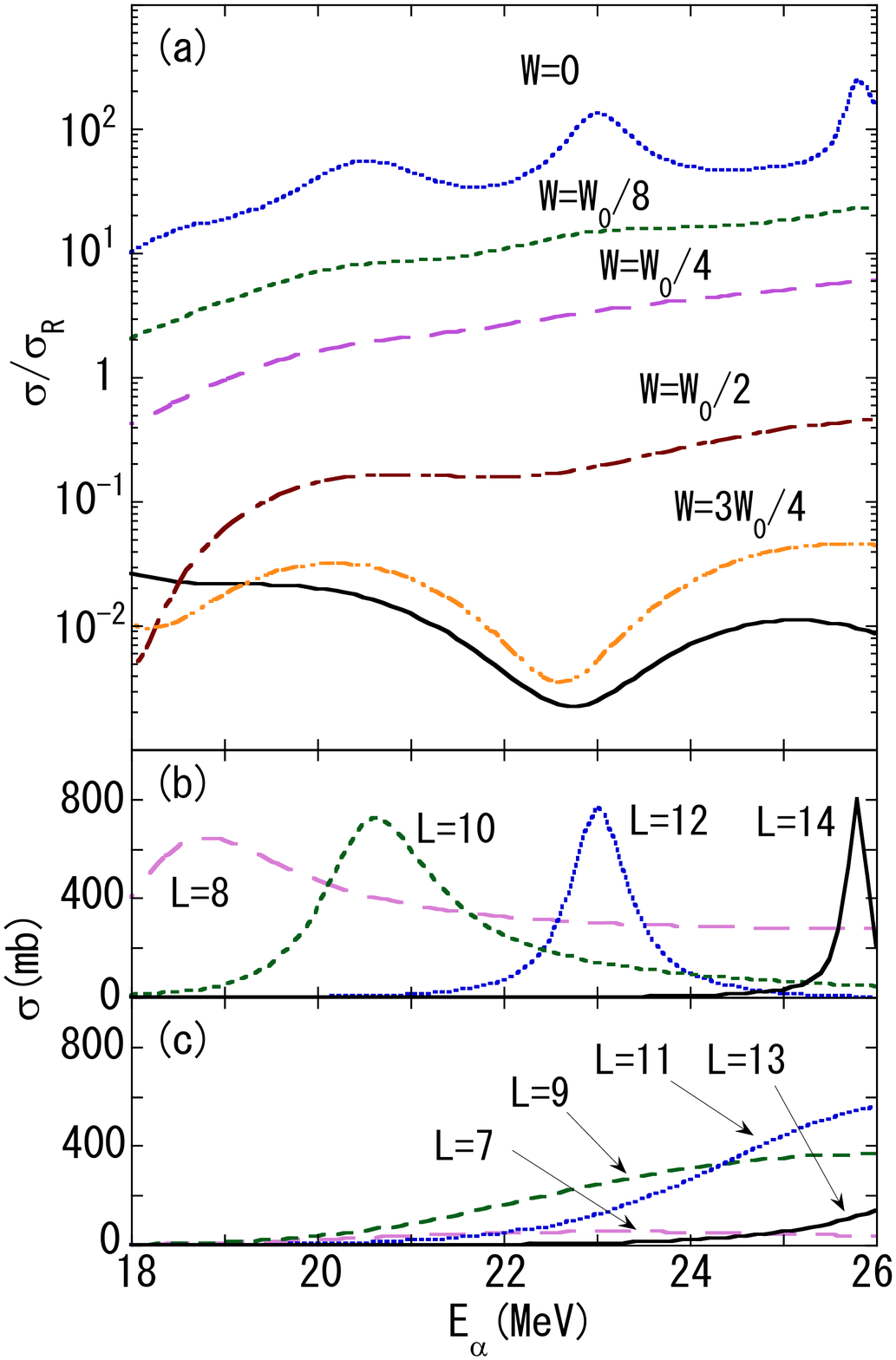}
\protect\caption {(Color online) (a)The excitation functions at the extreme backward angle $\theta$=176.2$^\circ$ in $\alpha$+$^{92}$Zr scattering calculated with the reduced strengths of the imaginary potential $W$=0 (dotted lines), $W_0$/8 (dashed lines), $W_0$/4 (long dashed lines), $W_0$/2 (dash-dotted lines), and 3$W_0$/4 ( dashed-and-double-dotted lines) are compared with the original one with $W=$$W_0$=10.7 MeV (solid lines).  (b) The calculated   partial wave cross sections of elastic scattering  under $W=$0 for  even $L$  and for (c) odd  $L$.}
\label{fig:ExfuncWdep}
\end{figure}

 For this purpose, in Fig.~\ref{fig:ExfuncWdep}~(a)  back-angle excitation functions calculated by reducing gradually the strength of the imaginary potential, $W$=3$W_0$/4, $W_0$/2, $W_0$/4, $W_0$/8, and 0 MeV, are compared with the original one with $W_0=10.7$  in Table II.
 For $W=0$ the peaks in the excitation function  at $E_\alpha$=20.5   and 23 MeV
are due to the resonances with the  $\alpha$+$^{92}$Zr structure. That the $\alpha$ cluster structure  at the highly excitation energies  can be seen in the  excitation function at the extreme backward angles near $180^\circ$ has been already shown for the $\alpha$+$^{40}$Ca cluster structure in $^{44}$Ti \cite{Michel1986B,Ohkubo1987}.

% Fig.4 W=0 W/4 W/2 S-matrix 1-S alpha+ 92Zr 

% Fig5 phase shift alpha+ 92Zr 
\begin{figure}[t!]
\includegraphics[width=8.6cm]{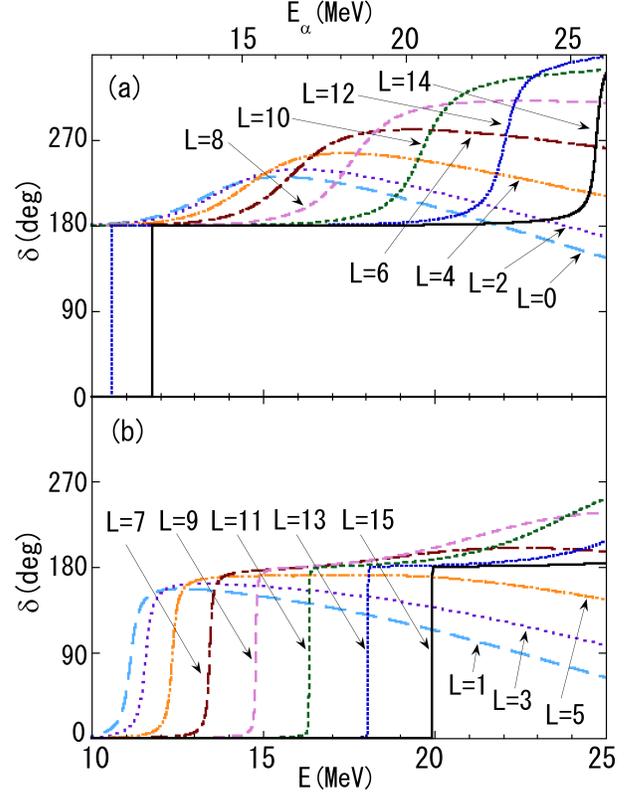}% Here is how to import EPS art
\protect\caption {(Color online) Phase shifts in $\alpha$  + $^{92}$Zr scattering  calculated with the double folding potential with  $N_R$=1.22,  (a) even $L$ partial waves   and (b) odd $L$ partial waves, are displayed for $L\le$15.
The lower abscissa shows cener-of-mass  energy $E$ and the upper abscissa shows laboratory incident energy $E_\alpha$. }
\label{fig:phaseshift}
\end{figure} 

 In Fig.~\ref{fig:ExfuncWdep}~(b) and (c),  the   partial wave cross sections of  elastic scattering are displayed. 
Fig.~\ref{fig:ExfuncWdep}~(b) shows that the peaks at  $E_\alpha$=20.5 and 23 MeV are caused by the even $L$ partial waves and Fig.~\ref{fig:ExfuncWdep}~(c) shows that the odd $L$ partial  waves do not contribute to create the peaks. Thus  we find  that the peaks at $E_\alpha$=20.5 and   $E_\alpha$=23 MeV in the excitation function with  $W=0$  are   caused by the resonant waves $L=10$  and $L=12$, respectively.

\par 
To see the details of the resonances responsible for the peaks, in Fig.~\ref{fig:phaseshift} the phase shifts in $\alpha$+$^{92}$Zr scattering calculated by switching off the imaginary potential are displayed.
We see that the phase shifts for the even parity and odd parity partial waves show  different behavior in the relevant energies, $E_\alpha$=18 - 26 MeV (center-of-mass energy $E$=17.3-24.9 MeV).
Although  the the phase shifts of the even-parity partial waves, $L=$10 and 12, pass through $\delta_L$=270$^\circ$ slowly at the resonance energies, those of the odd-parity partial waves, $L=$11 - 15, cross $\delta_L$=90$^\circ$ sharply at the resonance energies. 
The narrow odd-parity resonances hardly contribute to the peaks in the excitation functions as seen in Fig.~\ref{fig:ExfuncWdep}~(a) and (c). This is  why even-parity waves are dominantly responsible for the peaks and the dip in Fig.~\ref{fig:ExfuncWdep}. The broad resonant nature of the even- parity waves  is the consequence   that they are the high-lying second-higher nodal $\alpha$-cluster resonance states, in which the relative motion is two more excited compared with the lowest  Pauli-allowed  ground-band states in $^{96}$Mo. The nature of the resonant $\alpha$+$^{92}$Zr cluster structure in $^{96}$Mo is discussed in detail in the next section.

\section{ ALPHA CLUSTER STRUCTURE IN $^{96}$M\lowercase{o}
AND NEUTRINOLESS DOUBLE $\beta$ DECAY of $^{96}$Z\lowercase{r}}

\par
\begin{figure}[t!]
\includegraphics[width=8.6cm]{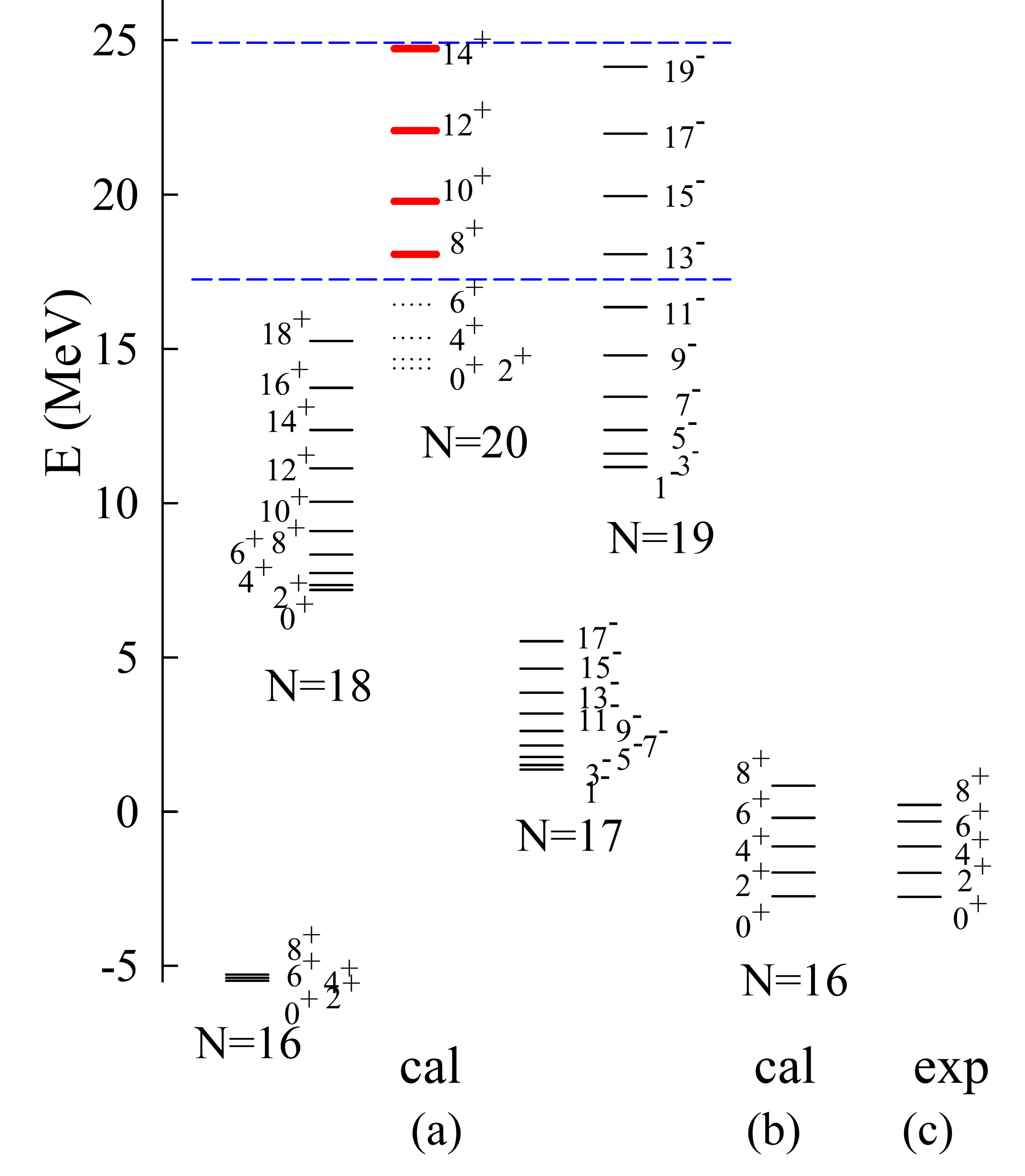}% Here is how to import EPS art
\protect\caption {(Color online)  (a) Energy levels of $^{96}$Mo calculated in the $\alpha$+ $^{92}$Zr cluster model with the double folding potential with
$N_R$=1.22.
The calculated excitation energy of the $N$=16 ground-band states, $0^+$, $2^+$, $4^+$,  $6^+$ and  $8^+$, which look   compressed, increases  as the spin increases.  
 (b) The   $N=16$ band energy levels calculated using  the  double folding model with $L$ dependence. (c) Experimental energy levels of the ground band.  The horizontal dashed lines (blue) correspond to $E_\alpha$=18 MeV (center-of-mass energy $E$=17.3) and $E_\alpha$=26 MeV ($E$=24.9 MeV), between  which the characteristic dip in the excitation function appear.
}
\label{fig:EnergyLevel}
\end{figure}
\begin{figure}[t!]
\includegraphics[width=8.0cm]{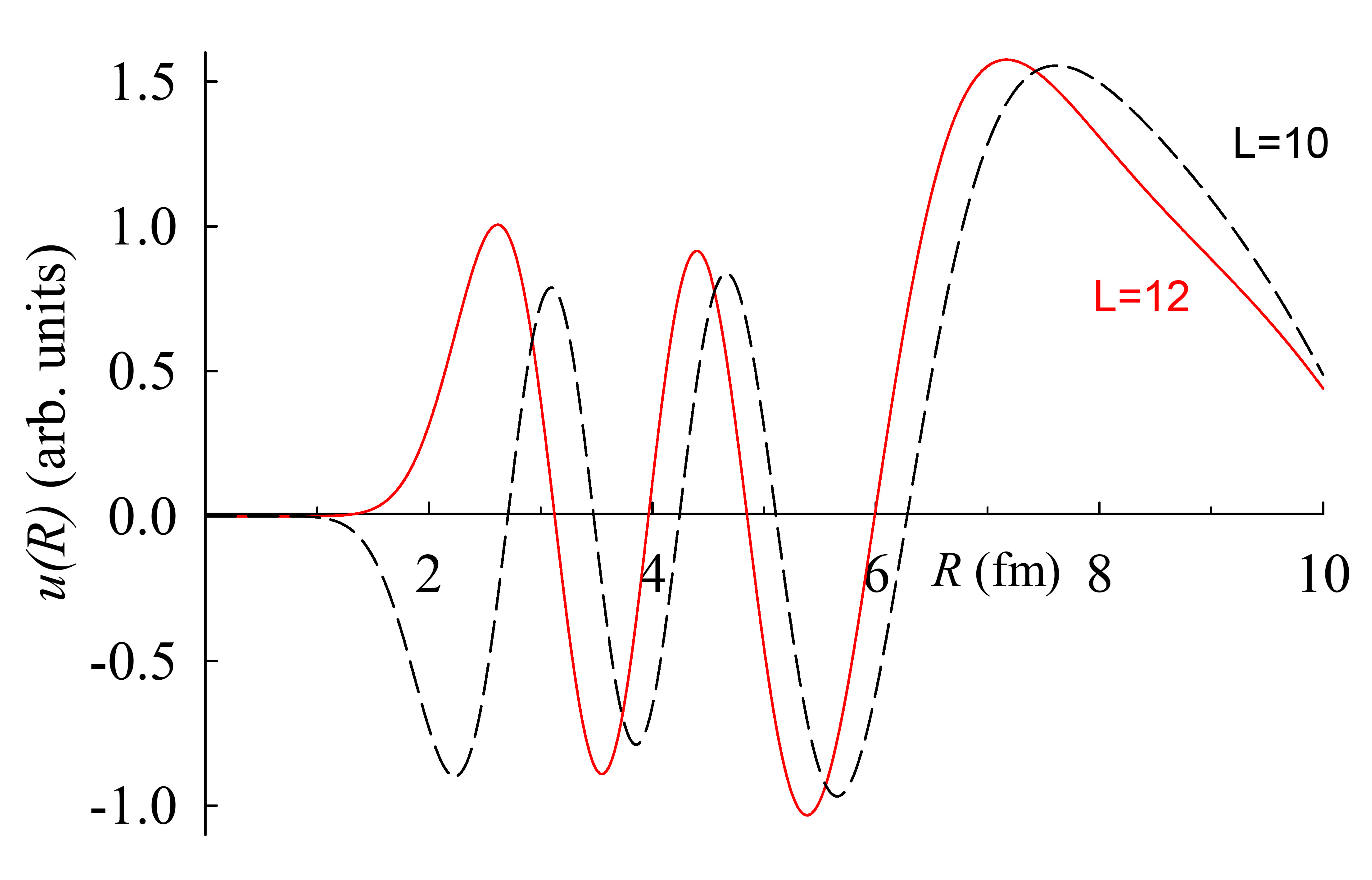}% Here is how 
\protect\caption{(Color online) The  calculated $u(R)$ of the relative wave function $u(R)$/$R$  of the  $10^+$ and $12^+$ states of  the  $N=$20 band with the $\alpha$+$^{92}$Zr cluster structure in $^{96}$Mo.  The wave functions are calculated as scattering states and normalized to unity for $R\le$10 fm.
}
\label{fig:N=20L=12L=10}
\end{figure}

In order to reveal the cluster structure of $^{96}$Mo underlying in the excitation function with the characteristic dip at the extreme backward angle, the resonant states  and the bound and quasibound energy levels calculated in the double folding potential with $N_R$=1.22 by switching off the imaginary potential of the optical potential used in Fig.~3 are displayed in Fig.~\ref{fig:EnergyLevel}~(a). The resonance energies are given at the energies where the phase shifts  steeply pass through $\delta_L$=90$^\circ (270^\circ)$ in Fig.~\ref{fig:phaseshift}.
By investigating the resonant wave function for $L=12$ at $E=22.04$ MeV, we find that the wave function has four nodes in the relative motion, see  Fig.~\ref{fig:N=20L=12L=10}.
The resonances with $L$=10, 12, and 14  in the range of $E$=19-25 MeV  are found to belong to the band with $N=2n+L=20$, where $n$ is the number of the nodes in the relative wave function between $\alpha$ and $^{92}$Zr. The $N=20$ band states energies are  on the $J(J+1)$ plot with the bandhead $J^\pi$=$0^+$ state at
$E$=14.4 MeV and the rotational constant $k$=$\hbar^2/2 \cal{J}$=0.0492 MeV, where $\cal{J}$ is the moment of inertia of the band. 
 The band has a well-developed $\alpha$+$^{92}$Zr cluster structure. The large separation distance between $\alpha$ and  $^{92}$Zr  can be seen 
in the the wave functions  of  the $10^+$ and $12^+$ states in Fig.~\ref{fig:N=20L=12L=10}. The  outermost peak, which  is located  at around  $R$=7-8 fm, is much larger than the sum of the experimental radius \cite{Angeli2013} of  $\alpha$ and  $^{92}$Zr, 6.0 fm.  Although the phase shifts for the lower $L=0-6$ of the $N=20$ band show rising toward 
$\delta_L$=270$^\circ$, they do not  cross  $\delta_L$=270$^\circ$ sufficiently. However, since the number of the nodes $n$ of their wave functions satisfy the condition $N=2n+L=20$, they are considered to belong to the persistent 
member of the rotational band with $N=20$. From the 
$J(J+1)$ plot they are extrapolated to exist persistently at the energies indicated by the dotted lines in Fig.~\ref{fig:EnergyLevel}~(a).
The resonance energies and widths of these broad resonances can be calculated in the complex scaling method \cite{Aguilar1971,Ohkubo2002}.
The presence of the $12^+$ state of the $N=20$ band, which manifests itself in the emergence of the characteristic dip in the back-angle excitation function, demonstrates for the first time the existence of a second-higher nodal band member state with the $\alpha$ + $^{92}$Zr cluster structure in $^{96}$Mo, in which two more nodes are excited in the relative motion compared with the $N=$16 ground band. 

% Fig6 energy levels of 96Mo N=20 band N=18,17 16 alpha+ 92Zr 

The wave functions of the resonances with odd $L$ in Fig.~\ref{fig:phaseshift} have   $N=19$  and form a negative-parity rotational band with the $\alpha$+$^{92}$Zr cluster structure. The band states are well located on the $J(J+1)$ plot with its bandhead $1^-$ state at $E=11.1$ MeV and $k$=0.0383 MeV. 
The  $N=19$ band is a higher nodal band with developed $\alpha$ clustering, in which the relative motion is one more excited compared with the lower-lying $N=17$ band states. 
The calculation locates the $N=18$ rotational band with its bandhead $0^+$ at $E$=7.19 MeV and $k$=0.0236 MeV,  which  is a higher nodal band with  one more  node in the wave functions compared with  those of the Pauli-allowed lowest $N=16$ band.
The $N=17$ rotational band states are well located on the 
$J(J+1)$ plot with its bandhead $1^-$ state at $E$=1.33 MeV.
 The calculation locates the band states with $N=16$ below the $\alpha$ threshold. It is surprising that the Pauli-allowed lowest $N=16$ band states satisfying the Wildermuth condition falls in good correspondence with the ground band of $^{96}$Mo. The calculated $0^+$ state of the $N=16$ band with $E$=-5.56 MeV is slightly overbound by 2.8 MeV compared with the experimental energy of the ground state with $E=$-2.76 MeV from the $\alpha$ threshold. This is because the potential determined at the highly excited energy region, $E_\alpha$=23 MeV, is straightforwardly applied to the calculations in the bound-state energy region.  The energy levels for $N=$18, 17 and 16 in Fig.~\ref{fig:EnergyLevel}, most of which are located below the Coulomb barrier, are the ones calculated in the bound-state approximation. 

%%% Fig. 8  N=20 L=12 wf

According to the dispersion relation \cite{Mahaux1986}, the energy dependence of the volume integral of the real potential shows the threshold anomaly. Namely the volume integral $J_V$  increases as the incident energy decreases from the rainbow energy region to the lower-energy region of BAA and  reaches  a maximum  followed by a decrease toward $E_\alpha$=0.
In fact, in Table I and II we see that $J_V$ increases
from 280.5 MeVfm$^3$ at the rainbow energy $E_\alpha$=120 MeV to 318.9 MeVfm$^3$ at $E_\alpha$=35.4 MeV and 329.6 MeVfm$^3$ at $E_\alpha$=23 MeV. 
The dispersion relation tells that a  potential with a reduced $J_V$ value should be used in the bound and quasibound energy region below and near the threshold energy $E$=0.
The overbinding of the ground-state energy in Fig.~\ref{fig:EnergyLevel}~(a) 
is simply ascribed to that it is calculated using the potential with $N_R$=1.22 at $E_\alpha$=23 MeV with the large $J_V$ value without taking into account the energy dependence of the real potential due to the dispersion relation \cite{Mahaux1986}.
By using the double folding potential with a slightly reduced strength, $N_R$=1.182 with $J_V$=319.3 MeVfm$^3$, the calculated ground-state $0^+$ energy agrees with the experimental value as seen in Fig.~\ref{fig:EnergyLevel}~(b).
A similar situation where $J_V$ must be reduced in the bound and quasibound energy region compared with that used in the higher scattering energy region has been reported in the recent unified description of bound and scattering states for the $\alpha$+$^{48}$Ca cluster structure in $^{52}$Ti \cite{Ohkubo2020} and the $\alpha$+$^{44}$Ca cluster structure in $^{48}$Ti \cite{Ohkubo2021}.

% Table III excitation energies removed 2021.7.23
\begin{table}[t!]
\protect\caption{ The excitation energies $E_x$, intercluster rms radii $\sqrt{<R^2>}$
and $B(E2)$ values for the $J\rightarrow (J-2)$ transitions of the  ground band in $^{96}$Mo calculated in the $\alpha$+$^{92}$Zr cluster model 
with the double folding potential are compared with the experimental data \cite{Abriola2008} and the large-scale shell model calculation \cite{Coraggio2018}. 
}
\begin{center}
\begin{tabular}{cllcccc}
\hline
\hline
$J^\pi$ & \multicolumn{2}{c}{$E_x$ (MeV)} & $\sqrt{<R^2>}$ (fm)& \multicolumn{3}{c}{$B(E2)$} (W.u.) \\
& exp & cal & cal &exp \cite{Abriola2008} &cal & Ref.\cite{Coraggio2018}\\
\hline
$0^+$ & 0.00 & 0.000 &5.20 & & & \\ 
$2^+$ & 0.778 & 0.770 &5.21 &20.7$\pm$0.4 & 20.7 &18.7 \\ 
$4^+$ &1.628 &1.611 & 5.19& 41$\pm$7 & 28.7 & - \\ 
$6^+$ &2.441 &2.543 & 5.14& - & 29.1 & - \\ 
$8^+$ &2.978 &3.583 & 5.05& - & 26.5 & - \\ 
\hline
\hline
\end{tabular}
\end{center}
\end{table}

\par
In Fig.~\ref{fig:EnergyLevel}~(a) the calculated $N$=16 ground-band states  are very compressed, which  is also the case for the $N=16$ states calculated with $N_R$=1.182. 
 Although the conventional Wood-Saxon potential gives an inverted energy level spectrum in this heavy mass region, namely the excitation energy  of the ground- band   states decreases as the spin increases in disagreement with experiment, the present double folding model potential  gives the  energy spectrum   consistent with  experimental ground band. In fact,  the  excitation energy   of the calculated energy levels of the $N$=16 band,  which  looks almost degenerate in Fig.~\ref{fig:EnergyLevel}~(a),   increases as the spin increases from  $0^+$ to   $8^+$.
This compression is because the angular momentum dependence of the local potential has not been taken into account.
In order to discuss the spectroscopic properties in the low-energy region, it is  necessary to take into account the $L$ dependence of the potential. 
 The nucleus-nucleus  potential, which is originally non-local due to the 
the Pauli principle, has $L$ dependence  when represented as a local potential. The $L$  dependence is usually not important and often neglected in the scattering energy region. However, this $L$ dependence 
is important when we study the cluster structure in the bound and quasibound  energy region.
The necessity of the $L$ dependence of the intercluster potential due to the Pauli principle has been theoretically founded in the  microscopic studies of interactions between composite particles
\cite{Tohsaki1980,Aoki1982}. In fact, it has been shown that this $L$ dependence is indispensable in 
the $\alpha$ cluster structure using a local potential, for example,
in $^{20}$Ne \cite{Michel1989}, $^{44}$Ti \cite{Michel1986,Michel1988}, 
$^{94}$Mo \cite{Ohkubo1995,Souza2015}, $^{212}$Po \cite{Ohkubo1995,Ni2011}, and $^{46,50}$Cr \cite{Mohr2017}. 
Following the double folding potential model study of the $\alpha$ cluster structure in $^{94}$Mo in Ref. \cite{Ohkubo1995} where linear  $L$ dependence in the double folding potential is first discussed, we use $N_R^{(L)}$=$N_R^{(L=0)}$ - $c$ $L$ with $N_R^{(L=0)}$=1.182 and $c$=5.00$\times10^{-3}$ for $^{96}$Mo.
The calculated energy levels of the $N=16$ ground band are displayed in Fig.~\ref{fig:EnergyLevel}~(b). 
In Table III the calculated $B(E2)$ values as well as the excitation energies, intercluster rms radii of the ground band of $^{96}$Mo are listed in comparison with the experimental data.
The excitation energy of the ground band is reproduced well by the double folding potential model with small $L$ dependence.
The experimental $B(E2)$ values \cite{Abriola2008} are also reproduced well by introducing an additional small effective charge $\Delta e=0.3 e$ for protons and neutrons.
We note that in the large-scale shell-model calculations in Ref. \cite{Coraggio2018} 
rather large additional effective charges $\Delta e=0.5 e$ for protons and $\Delta e=0.5-0.8 e$ for neutrons are introduced.
The rms charge radius $<r^2>_{^{96}{\rm Mo}}^{1/2}$=4.36 fm of the ground state
calculated using the experimental values $<r^2>_{^4{\rm He}}^{1/2}$=1.676 fm and $<r^2>_{\rm ^{92}Zr}^{1/2}$=4.306 fm \cite{Angeli2013} is in good agreement 
with the experimental value 4.38 fm \cite{Angeli2013}. The calculated intercluster distance of the ground state is about 87\% of the sum of the experimental rms charge radii of the two clusters, which is compared to 87\% for the ground state of $^{44}$Ti \cite{Michel1988}. 

\par
 Note that the value of the parameter $c$=5.00$\times10^{-3}$ lies in the expected range of $\alpha$-cluster states $c$$\approx$(2.5-5)$\times 10^{-3}$, as observed for many $\alpha$-cluster states in a wide range of nuclei \cite{Mohr2007} including the mass region near $A=100$ such as $^{94}$Mo \cite{Ohkubo1995,Michel1998}, $^{93}$Nb \cite{Ohkubo2009,Kiss2009}, and $^{104}$Te \cite{Mohr2007}; light- and medium-weight mass regions such as $^{20}$Ne \cite{Abele1993} and $^{44}$Ti \cite{Atzrott1996}, and heavy mass region such as $^{212}$Po \cite{Hoyler1994,Ohkubo1995}.
 The $L$-dependent potential calculation locates the negative-parity $N=17$ band with the head $1^-$ state at $E_x$=6.83 MeV well below the Coulomb barrier. 
Although $1^-$ states have been observed at $E_x$=3.600 MeV and 3.895 MeV \cite{Abriola2008}, experimental spectroscopic properties about $\alpha$ clustering of the excited states near and above $E_x$$\approx$4 MeV are not clear.

%%% Fig.8 96Mo gs wave function
\begin{figure}[t!]
\includegraphics[width=8.0cm]{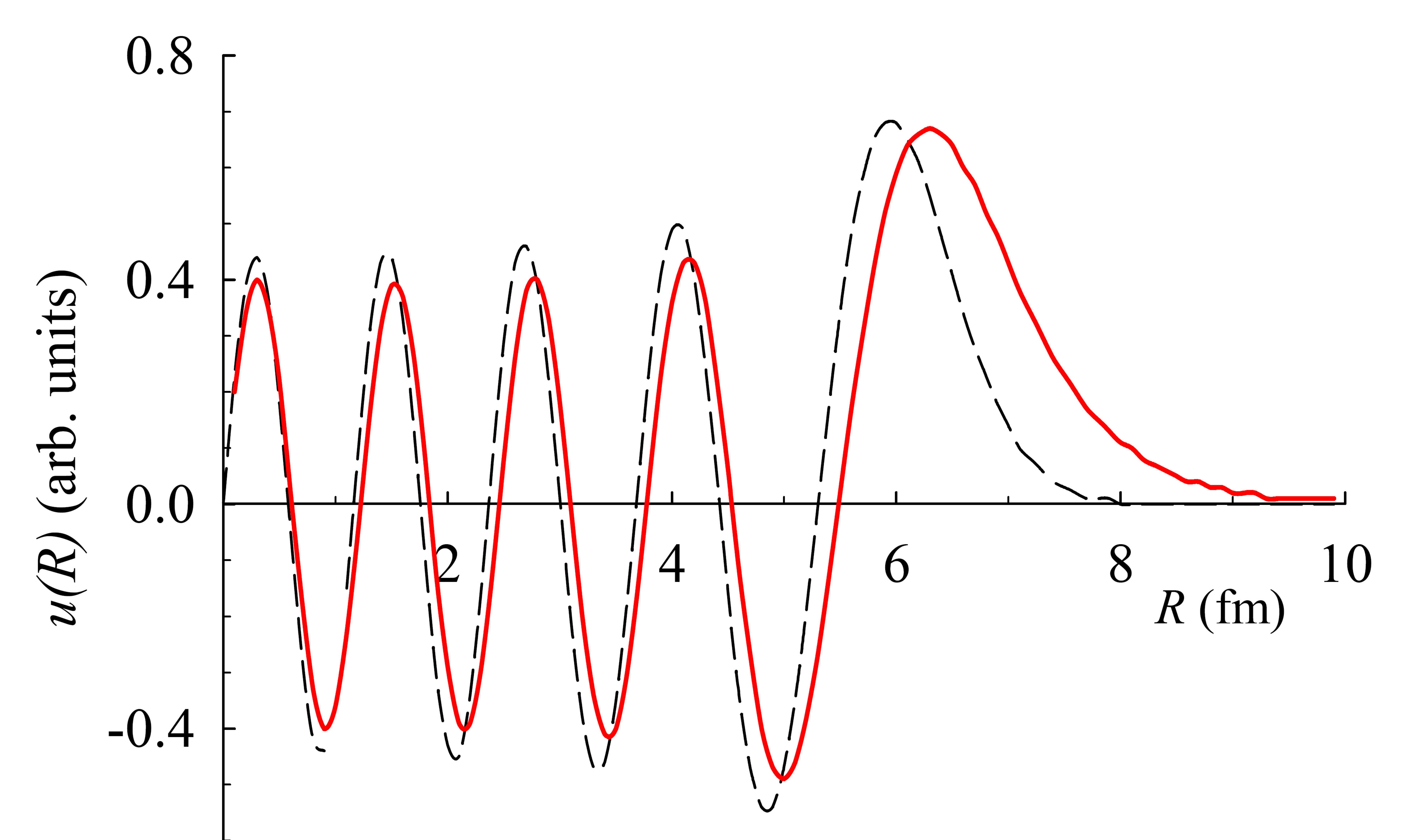}% Here is how 
\protect\caption{(Color online) The calculated $u(R)$ of the relative wave function $u(R)$/$R$ of the ground state of $^{96}$Mo calculated in the $\alpha$+ $^{92}$Zr cluster model (solid line) is compared with the harmonic oscillator wave function with $N_{\rm HO}$ =16 (dashed line). 
}
\label{fig:Mo96wf}
\end{figure}

\par 
The potential embeds the eight deeply bound unphysical Pauli-forbidden $0^+$ states. 
The overlaps of the eight $0^+$ wave functions  with the harmonic oscillator wave functions with $\nu$=0.22 fm$^{-2}$ ($\nu=m\omega/\hbar$ and $\hbar\omega$=9.12 MeV) are 0.96, 0.93, 0.93, 0.95, 0.97 and 0.98, 0.99, and 0.97 for the oscillator quanta $N_{\rm HO}$=$0$, $2$, $\cdots$, and $14$, respectively. This means that the Pauli-forbidden states with $N_{\rm HO}<16$ in the resonating group method are highly excluded from the obtained ground-state wave function, thus mimicking Saito's orthogonality condition model \cite{Saito1968}. 

\par
As seen in Fig.~\ref{fig:Mo96wf}, the ground-state wave function resembles the shell-model function with $N_{\rm HO}=16$
in the internal region. However, the outermost peak at around 6 fm is slightly shifted outward compared with that of the harmonic oscillator wave function causing significant
enhancement of the amplitude at the outer surface region due to $\alpha$ clustering. This enhancement means that the obtained wave function contains significant amount of components in the shells higher than 
$N_{\rm HO}$ =16.

\par
\begin{figure}[t!]
\includegraphics[width=8.0cm]{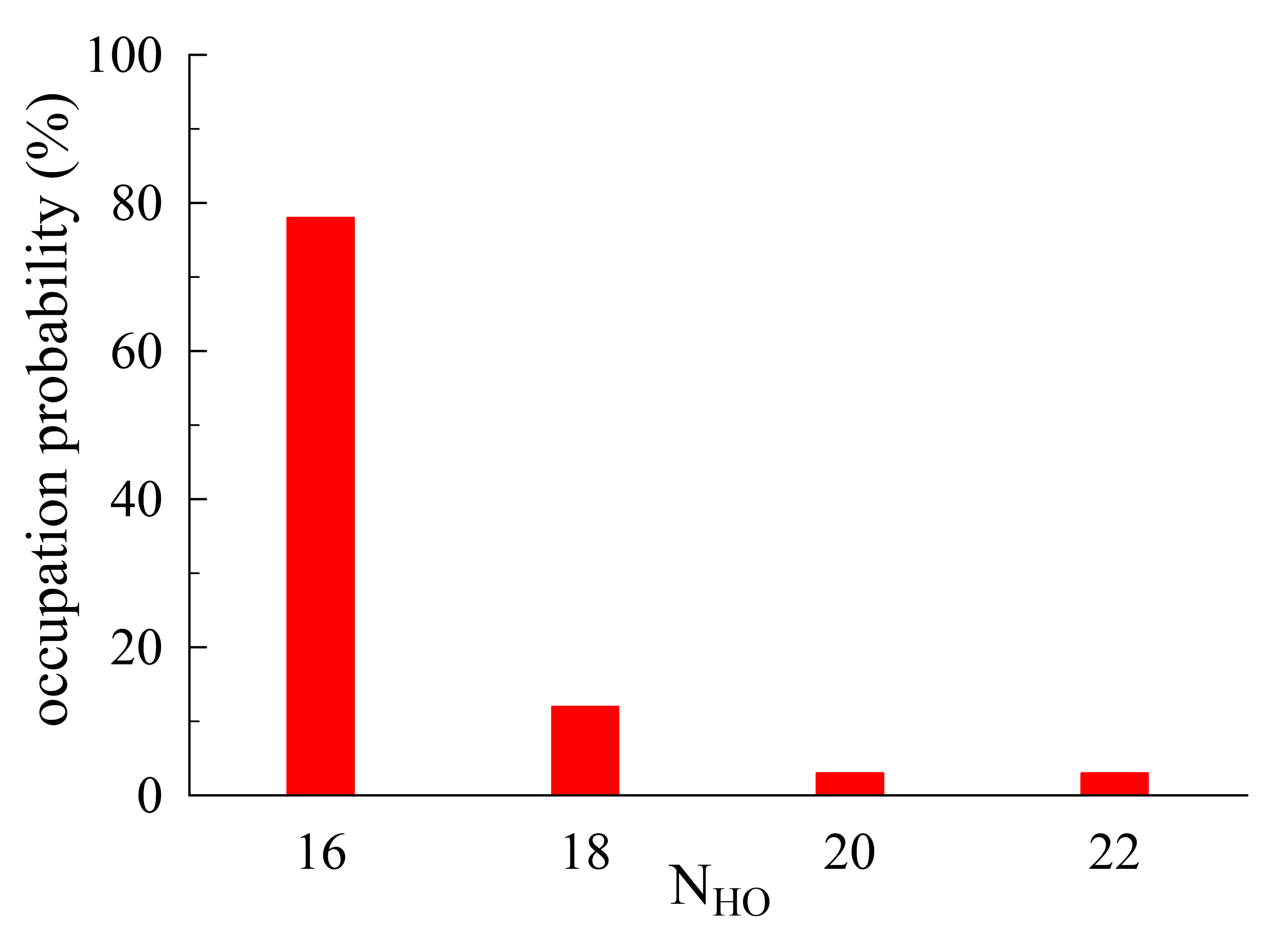}% Here is how 
\protect\caption{(Color online) The occupation probability of the harmonic oscillator quanta $N_{\rm HO}$ in the ground-state wave function of $^{96}$Mo. }
\label{fig:Nquanta}
\end{figure}

In Fig.~\ref{fig:Nquanta} the occupation probability of the quanta $N_{\rm HO}\ge$16 in the ground state wave function is displayed. 
The dominant occupation probability in the lowest Pauli-allowed shell-model like $N_{\rm HO}=16$ is 78\%. 
The  significant amount of higher $N_{\rm HO}\geq18$ components, 22\% 
is due to the $\alpha$ clustering of the ground state. 
The $2^+$ and $4^+$ states have the similar character. This $\alpha$ clustering
is responsible for the enhancement of the $B(E2)$ values in $^{96}$Mo and should be 
also taken into account in the
evaluation of NME of $0\nu\beta\beta$ decay of $^{96}$Zr to $^{96}$Mo. 

%%% Fig. 9 N quanta probability

\par
We discuss the effect of $\alpha$ clustering of $^{96}$Mo on the NME of 
$0\nu\beta\beta$ decay of $^{96}$Zr, which was the motivation of the present $\alpha$ cluster structure study of $^{96}$Mo as introduced in Sec. I.
The NME values of $0\nu\beta\beta$ decay of $^{96}$Zr evaluated by using various nuclear models are summarized in Ref. \cite{Engel2017} 
and most recently in Ref. \cite{Nautiyal2022}. 
The QRPA calculations give NME values  2.72 with the Argonne V18 nucleon-nucleon potential and 2.96 with the CD-Bonn nucleon-nucleon potential with the axial vector coupling constant $g_A$=1.27 in Ref. \cite{Simkovic2013} and 3.14 with $g_A$=1.26 in
 Ref. \cite{Hyvarinen2015}.  The IBM calculation by Barea {\it et al.} gives 2.53 in Ref. \cite{Barea2013}. 
The latest PHFB calculations give 2.5 \cite{Nautiyal2022}.
On the other hand, the EDF calculations give considerably large NME values,
about twice as large as the above values. The nonrelativistic EDF calculations by Vaquero {\it et al.} \cite{Vaquero2013} give 5.65 with $g_A$=1.26 when evaluated with shape fluctuation and 6.50 when evaluated with both the shape and pairing fluctuations. The relativistic EDF calculation by Yao {\it et al.} \cite{Yao2015} gives almost the same large result.
Yao {\it et al.} claim \cite{Yao2015} that the EDF calculations are unable to reproduce the properties of $^{96}$Zr giving too-low excitation energy of E$(2_1^+)$ and too-large $B(E2: 0_{\rm g.s.} \rightarrow 2_1^+)$ value, which is one order of magnitude large  compared with the experimental data.
Yao {\it et al.} \cite{Yao2015} ascribe this to the overestimation of the collectivity in $^{96}$Zr due to the 
``common problem of most EDF-based GCM or collective
Hamiltonian calculations.'' 
 Moreover, the GCM calculation in the frame of covariant density functional theory \cite{Song2017} gives the largest value of 6.37 among the nuclear model calculations. 
The overestimation of the collectivity of the doubly closed-subshell nucleus $^{96}$Zr increases the overlap of the wave functions of $^{96}$Zr and $^{96}$Mo, which leads to the large NME values. 
Although the present cluster model is unable to calculate NME values because nucleon degree of freedom is not involved, it can qualitatively tell whether the NME is enhanced or reduced by $\alpha$ clustering of $^{96}$Mo compared with the shell-model calculations in which the excitations to the higher major shells are not included.
 Taking into account that the excitation energy 1.58 MeV of the first excited state $0^+$ of $^{96}$Zr is rather high in this mass region resembling the mysterious $0^+$ of the double magic nucleus $^{16}$O \cite{Molnar1986}, and that there is no evidence that $^{96}$Zr has $\alpha$+$^{92}$Sr clustering \cite{Molnar1986}, the ground-state wave function can be well considered to have a doubly closed-subshell shell-model structure.  Thus the $\alpha$ clustering of $^{96}$Mo reduces considerably the overlap of the ground-state wave function of $^{96}$Zr with that of $^{96}$Mo in the evaluation of the NME. That is,
the $0\nu\beta\beta$ decay of $^{96}$Zr to $^{96}$Mo would be significantly quenched, thus have a longer half-life, due to the $\alpha$ clustering than that in the shell-model calculations which do not take into account the four particle excitations. 
Unfortunately, NME values in the shell model have not been reported. 
The shell-model calculations with configuration mixing including $N_{\rm HO}=$ 18, 20 and 22 major shells are presently formidably difficult even with  modern computers.
We note that both the QRAP and IBM calculations do not include $\alpha$-like four-particle four-hole excitations and $\alpha$-like correlations.

\par
Finally, we briefly mention about
the  large $B(E2)$ value 51.7 W.u. of the  transition from  $0_2^+$  ($E_x$=1.148 MeV)  to  $2^+_1$  ($E_x$=0.778 MeV)  in $^{96}$Mo  \cite{Abriola2008}, which  may  suggest that  the $0_2^+$ state  has  $\alpha$ clustering in which  the  core is  excited.  
  If the  $0_2^+$  state has significant amount of $[\alpha_{(L=2)}$+$^{92}$Zr(2$^+_1)]_{J=0}$ clustering component, then  the $B(E2)$ value  can be enhanced   because  in addition to the $E2$ transition matrix element due to the intercluster relative motion, <2$^+_1(\alpha_{L=0})|\hat{O}_{E2}( {\bf r})| 0_2^+(\alpha_{L=2}$)>, the  internal  transition  of the core  $^{92}$Zr, <$^{92}$Zr(g.s.) $|\hat{O}_{E2} (\xi)|^{92}$Zr($2^+_1$)>,    contributes to the total $E2$ transition   where  $\xi$ is the internal coordinate of $^{92}$Zr. Coupled- channels calculations with  excitations of  $^{92}$Zr would  be a future challenge  to understand the origin of the large $B(E2)$ value of the $0_2^+$ state of $^{96}$Mo and the effective charge.

\section{SUMMARY}
\par
In the evaluation of nuclear matrix element of neutrinoless double $\beta$ decay $0\nu\beta\beta$ of the doubly closed-subshell nucleus $^{96}$Zr to $^{96}$Mo, it is important to take into account the collectivity due to $\alpha$ clustering in the structure of $^{96}$Mo, 
which has extra two neutrons on the $^{94}$Mo nucleus, which is analog of $^{20}$Ne and $^{44}$Ti and has been considered to have $\alpha$ cluster structure. 
We have studied for the first time $\alpha$ clustering aspects of $^{96}$Mo by using a double folding potential determined from the analysis of nuclear rainbows at high energies and the  characteristic structure of the angular distributions at low energies in $\alpha$ particle scattering from $^{92}$Zr.
The validity of the double folding potential used is also confirmed by studying $\alpha$ scattering from $^{90,91.94}$Zr in the low-energy region where a characteristic  dip appears in the excitation functions at the extreme backward angle near $180^\circ$.
The double folding model calculations reproduced  well all the observed angular distributions over a wide range of incident energies and the  excitation functions with a characteristic dip at the extreme backward angle.
 By studying the $\alpha$ cluster structure with the obtained double folding potential, the existence of the second-higher nodal $N=20$ band states with the $\alpha$+ $^{92}$Zr cluster structure, in which two more nodes are excited in the relative motion compared with the $N=16$ ground band in $^{96}$Mo, is demonstrated for the first time at the highly excited energy region. 
The $\alpha$-cluster model using this potential locates the ground state in agreement with experiment and reproduces the observed $B(E2)$ value of $^{96}$Mo. 
The effect of $\alpha$ clustering in $^{96}$Mo on the the half-life of the $0\nu\beta\beta$ double-$\beta$ decay of $^{96}$Zr is discussed. 

\par
\begin{acknowledgements}
One of the authors (S.O.) thanks the Yukawa Institute for Theoretical Physics, Kyoto University where part of the work was done during a stay in 2022.
\end{acknowledgements}


\begin{thebibliography}{reference}
%%%%%%%%%%%
\bibitem{Ejiri2005} H. Ejiri, J. Phys. Soc. Jpn. {\bf 74}, 2101 (2005).
\bibitem{Avignone2008} %half-life
F. T. Avignone, III, S. R. Elliott, and J. Engel, Rev. Mod. Phys.
{\bf 80}, 481 (2008). 
\bibitem{Vergados2012}%Theory of neutrinoless double-beta decay
J. D. Vergados, H. Ejiri, and F. $\breve{\rm S}$imkovic,
Rep. Prog. Phys. {\bf 75}, 106301 (2012).

\bibitem{Suhonen1998} 
J. Suhonen and O. Civitarese, Phys. Rep. {\bf 300}, 123 (1998).
\bibitem{Faessler1998}
A. Faessler and F. Simkovic, J. Phys. G: Nucl. Part. Phys. {\bf 24},
2139 (1998).
\bibitem{Suhonen2012}
J. Suhonen and O. Civitarese, J. Phys. G: Nucl. Part. Phys. {\bf 39},
085105 (2012).
%IBM

\bibitem{Engel2017}%Status and future of nuclear matrix elements for neutrinoless double-beta decay:
J. Engel and J. Men\'{e}ndez,
Rep. Prog. Phys. {\bf 80}, 046301 (2017).
\bibitem{Jokiniemi2018}
L. Jokiniemi, H. Ejiri, D. Frekers, and J. Suhonen, Phys. Rev. C
{\bf 98}, 024608 (2018).
\bibitem{Dolinski2019}%neutrinoless double beta decay: status and prospect
M. J. Dolinski, A. W. P. Poon, and W. Rodejohann,
Ann. Rev. Nucl. Part. Sci. {\bf 69}, 219 (2019).

\bibitem{Iwata2016}%Large-Scale Shell-Model Analysis of the Neutrinoless beta beta Decay of 48Ca
Y. Iwata, N. Shimizu, T. Otsuka, Y. Utsuno, J. Men\'{e}ndez, M. Honma, and T. Abe,
Phys. Rev. Lett. {\bf 116}, 112502 (2016).
\bibitem{Coraggio2020}%Calculation of the neutrinoless double- decay matrix element within the realistic shell model,
L. Coraggio, A. Gargano, N. Itaco, R. Mancino, and
F. Nowacki, Phys. Rev. C {\bf 101}, 044315 (2020).
\bibitem{Yao2020} %Ab initio treatment of collective correlations and the neutrinoless double beta decay of 48Ca,
J. M. Yao, B. Bally, J. Engel, R. Wirth, T. R. Rodr\'{i}guez,
and H. Hergert,
Phys. Rev. Lett. {\bf 124}, 232501 (2020).
\bibitem{Belley2021}%ab Initio Neutrinoless Double-Beta Decay Matrix Elements for 48Ca, 76Ge, and 82Se
A. Belley, C. G. Payne, S. R. Stroberg, T. Miyagi, and J. D. Holt,
Phys. Rev. Lett. {\bf 126}, 042502 (2021).
\bibitem{Simkovic2013} %96Zr QRPA
F. \u{S}imkovic, V. Rodin, A. Faessler, and P. Vogel,
Phys. Rev. C {\bf 87}, 045501 (2013).

\bibitem{Hyvarinen2015} %96Zr QRPA
J. Hyv$\ddot{\rm a}$rinen and J. Suhonen, Phys. Rev. C {\bf 91}, 024613 (2015).

\bibitem{Jokiniemi2021} % QRPA and shell model 
L. Jokiniemi, P. Soriano, and J. Men\'{e}ndez,
Phys. Lett. {\bf B 823}, 136720 (2021). 


% Projected HF 
\bibitem{Chaturvedi2008} %Nuclear deformation and neutrinoless double-beta decay of 94,96Zr, 98,100Mo
K. Chaturvedi, R. Chandra, P. K. Rath, P. K. Raina, and J. G. Hirsch,
Phys. Rev. C {\bf 78}, 054302 (2008).
\bibitem{Rath2010} %96Zr 
P. K. Rath, R. Chandra, K. Chaturvedi, P. K. Raina, and
J. G. Hirsch, Phys. Rev. C {\bf 82}, 064310 (2010);
P. K. Rath, R. Chandra, K. Chaturvedi, P. Lohani, and P. K. Raina,
Phys. Rev. C {\bf 93}, 024314 (2016).
\bibitem{Nautiyal2022}
V. K. Nautiyal, R. Gautam, N. Das, R. Chandra, P. K. Raina, and P. K. Rath,
Eur. Phys. J. A {\bf58}, 28 (2022). 
% GCM %48Ti
\bibitem{Rodriguez2010} %96Zr 4848Ca
T. R. Rodr\'{i}guez and G. Martinez-Pinedo, Phys. Rev.
Lett. {\bf 105}, 252503 (2010).
\bibitem{Hinohara2014} %76Ge
N. Hinohara and J. Engel,
Phys. Rev. C {\bf 90}, 031301(R) (2014).
\bibitem{Song2017}%beyond mean field covariant density functional GCM
S. Song, J. M. Yao, P. Ring, and J. Meng, Phys. Rev. C {\bf 95}, 024305 (2017).

\bibitem{Jiao2019} %Union of rotational and vibrational modes in generator-coordinate-type calculations with application to neutrinoless double-beta decay Xe
C. Jiao and C. W. Johnson, 
Phys. Rev. C {\bf 100}, 031303(R) (2019).

%density functional
\bibitem{Vaquero2013}
N. L. Vaquero, T. R. Rodr\'{i}guez, and J. L. Egido, Phys. Rev.
Lett. {\bf 111}, 142501 (2013).
\bibitem{Yao2015} % include 96Zr
J. M. Yao, L. S. Song, K. Hagino, P. Ring, and J. Meng,
Phys. Rev. C {\bf 91}, 024316 (2015).

\bibitem{Barea2009} %include 96Zr 
J. Barea and F. Iachello, Phys. Rev. C {\bf 79}, 044301 (2009);
J. Barea, J. Kotila, and F. Iachello, Phys. Rev. Lett. {\bf 109},
042501 (2012).
\bibitem{Barea2013}
J. Barea, J. Kotila, and F. Iachello, Phys. Rev. C {\bf 87}, 014315
(2013);
F. F. Deppisch, L. Graf, F. Iachello, and J. Kotila,
Phys. Rev. D {\bf 102}, 095016 (2020).
\bibitem {Ohkubo2021} %48Ti=alpha+44Ca
S. Ohkubo,
Phys. Rev. C {\bf 104}, 054310 (2021). 

\bibitem {Suppl1972}%Suppl 1972
K. Ikeda {\it et al.}, 
Prog. Theor. Phys. Suppl. No. {\bf 52}, 1 (1972) and references therein.
\bibitem {Suppl1980}%Suppl 1980
K. Ikeda {\it et al.}, 
Prog. Theor. Phys. Suppl. No. {\bf 68}, 1 (1980) and references therein.
\bibitem{Michel1998}
F. Michel, S. Ohkubo, and G. Reidemeister, Prog. Theor. Phys.
Suppl. {\bf 132}, 7 (1998).
\bibitem{Yamaya1998}
T. Yamaya, K. Katori, M. Fujiwara, S. Kato, and S. Ohkubo,
Prog. Theor. Phys. Suppl. {\bf 132}, 73 (1998).
\bibitem{Sakuda1998}
T. Sakuda and S. Ohkubo, Prog. Theor. Phys. Suppl. {\bf 132}, 103
(1998).
\bibitem {Ohkubo1999}
S. Ohkubo, T. Yamaya, and P. E. Hodgson,
Nuclear clusters, in {\it Nucleon-Hadron Many-Body
Systems}, (edited by H. Ejiri and H. Toki) 
(Oxford University Press, Oxford, 1999), p. 150 
and references therein.
\bibitem{Fukada2009}% 44Ti 7- 46Ti many states observed, 52Ti exp
M. Fukada, M. K. Takimoto, K. Ogino, and S. Ohkubo, Phys.
Rev. C {\bf 80}, 064613 (2009). 
\bibitem {Ohkubo2020} %52Ti=alpha+48Ca
S. Ohkubo,
Phys. Rev. C {\bf 101}, 041301(R) (2020).
\bibitem{Bailey2019}%%Extracting the spectral signature of alpha clustering in 44Ti,48Ti,52Ti using a continuous wavelet transform 
S. Bailey,% {\it et al.}, 
T. Kokalova, M. Freer, C. Wheldon, R. Smith, J. Walshe {\it et al.},
Phys. Rev. C {\bf 100}, 051302(R) (2019).
\bibitem{Bailey2021}% 52Ti wavelet 
S. Bailey, % {\it et al.}, 
T. Kokalova, M. Freer, C. Wheldon, R. Smith, J. Walshe
{\it et al.},
Eur. Phys. J. A {\bf 57}, 108 (2021). 

\bibitem{Fukuda2020}% ZICOS - Neutrinoless Double Beta Decay experiment using Zr-96 with an organic liquid scintillation -
Y. Fukuda {\it et al.},
J. Phys. Conf. Series {\bf 1468}, 012139 (2020).

\bibitem {Ohkubo1995}% alpha+90Zr , alpha+208Pb
S. Ohkubo, Phys. Rev. Lett. {\bf 74}, 2176 (1995).
\bibitem {Michel2000}%94Mo
F. Michel, G. Reidemeister, S. Ohkubo, 
Phys. Rev. C {\bf 61}, 041601(R) (2000).
\bibitem{Buck1995}
B. Buck, A. C. Merchant, and S. M. Perez
Phys. Rev. C {\bf 51}, 559 (1995).
\bibitem{Ohkubo2009}
S. Ohkubo,
Int. J. Mod. Phys. A {\bf 24}, 2035 (2009).
\bibitem{Souza2015} %-cluster structure in even-even nuclei around 94Mo
M. A. Souza and H. Miyake,
Phys. Rev. C {\bf 91}, 034320 (2015).
\bibitem{Ni2011}%alpha-cluster structure above doubly closed shells in a generalized density-dependent cluster model
D. Ni and Z. Ren,
Phys. Rev. C {\bf 83}, 014310 (2011).

\bibitem{Tanaka2021} 
J. Tanaka, Z. Yang, S. Typel, S. Adachi, T. Aumann {\it et al.}, Science {\bf 371}, 260 (2021).
\bibitem {Molnar1986}
G. Mol\'{n}ar, S. W. Yates, and R. A. Meyer,
Phys. Rev. C {\bf 33}, 1843(R) (1986).
\bibitem {Togashi2016}% Zr isotopes phase transition
T. Togashi, Y. Tsunoda, T. Otsuka, and N. Shimizu,
Phys. Rev. Lett. {\bf 117}, 172502 (2016).
\bibitem{Stoica1995}S. Stoica,
Phys. Lett. {\bf B 350}, 152 (1995).% QRPA BE2 of 96Zr Two-neutrino double-beta decay half-live of 96Zr and Mo to excited states of 96Mo
\bibitem {Lesher2007}%Low-spin structure of 96Mo studied with the (n, n gamma) reaction
S. R. Lesher {\it et al.},
Phys. Rev. C {\bf 75}, 034318 (2007).
\bibitem {Gregor2017}
E. T. Gregor {\it et al.}, %Shell evolution of stable N = 50-56 Zr and Mo
Eur. Phys. J. A {\bf 53}, 50 (2017).
\bibitem {Coraggio2018}
L. Coraggio, A. Gargano, and N. Itaco,
{\it Proceedings of the International Conference on} {\it Nuclear Theory in the Supercomputing Era 2016 (NTSE-2016)},  
edited by A. M. Shirokov and A. I. Mazur  (Pacific National University, Khabarovsk,
Russia, 2018), p. 226.
\bibitem{DeVries1987} % 13C charge density
H. De Vries, C. W. De Jager, and C. De Vries,
At. Data and Nucl. Data Tables {\bf36}, 495 (1987);
F. J. Kline, H. Crannell, J. T. O'Brien, J. McCarthy, and R. R. Whitney,
Nucl. Phys. {\bf A 209}, 381 (1973).

\bibitem {Kobos1982}
A. M. Kobos {\it et al.},
Nucl. Phys. {\bf A384}, 65 (1982);
A. M. Kobos {\it et al.},
Nucl. Phys. {\bf A425}, 205 (1984).
\bibitem {Satchler1979} 
G.~R.~Satchler and W.~G.~Love, Phys. Rep. {\bf 55}, 183 (1979).

\bibitem {Brandan1997} 
M.~E.~Brandan and G.~R.~Satchler, Phys. Rep. {\bf 285}, 143 (1997).
%\bibitem {Ohkubo2021} %48Ti S. Ohkubo, Phys. Rev. C {\bf 104}, 054310 (2021).
\bibitem {Michel1986}
F. Michel, G. Reidemeister, and S. Ohkubo, 
Phys. Rev. Lett. {\bf 57}, 1215 (1986).
\bibitem {Michel1988}
F. Michel, G. Reidemeister, and S. Ohkubo,
Phys. Rev. C {\bf 37}, 292 (1988).
\bibitem {Put1977}
L. W. Put and A. M. J. Paans,
Nucl. Phys. {\bf A 291}, 93 (1977).
\bibitem {Singh1986} %35.4 MeV data
P. Singh. D. Rychel, R. Gyufko, B. Van Kr\"{u}chten, M. Lahanas, and C. A. Wiedner,
Nucl. Phys. {\bf A 458}, 1 (1986).
%B. J. Lund, N. P. T. Bateman, S. Utku, D. J. Horen, and G. R. Satchler, Phys. Rev. {\bf 51}, 635 (1995).
\bibitem {Fuller1975}
R. C. Fuller, Phys. Rev.  C {\bf 12}, 1561 (1975);
K. W. McVoy and G. R. Satchler, Nucl. Phys.  {\bf A 417}, 157 (1984).

\bibitem {Ohkubo1977}
S. Ohkubo, Y. Kondo, and S. Nagata, 
Prog. Theor. Phys. {\bf 57}, 82 (1977). 
\bibitem{Michel1983}
F. Michel, J. Albinski, P. Belery, Th. Delbar, Gh. Gr\'{e}goire, B. Tasiaux, and G. Reidemeister,
Phys. Rev. C {\bf 28}, 1904 (1983).
\bibitem{Delbar1978} %alpha+44Ca pot
Th. Delbar,
Gh. Gr\'{e}goire, G. Pai\'{c}, R. Ceuleneer, 
F. Michel, R. Vanderpoorten {\it et al.},
Phys. Rev. C {\bf 18}, 1237 (1978).
\bibitem {Brink1985}
D. M. Brink, {\it Semi-Classical Methods for Nucleus-Nucleus Scattering}
(Cambridge University Press, Cambridge, UK, 1985).
\bibitem {Wit1975}%Back-angle elastic alpha scattering from Y and 90, 91,94Zr
M. Wit, J. Schiele, K. A. Eberhard, and J. P. Schiffer,
Phys. Rev. C {\bf 12}, 1447 (1975).
\bibitem {Michel1986B}
F. Michel, G. Reidemeister, and S. Ohkubo,
Phys. Rev. C {\bf 34}, 1248 (1986).
\bibitem {Ohkubo1987}
S. Ohkubo,
Phys. Rev. C {\bf 36}, 551 (1987).
\bibitem{Angeli2013}%Table of experimental nuclear ground state charge radii: An 
I. Angeli and K. P. Marinova,
At. Data and Nucl. Data Tables {\bf 99}, 69 (2013).
\bibitem {Aguilar1971}
J. Aguilar and J. M. Combes, Commun. Math. Phys. {\bf 22}, 269 (1971);
E. Balslev and J. M. Combes, Commun. Math. Phys. {\bf 22}, 280 (1971); 
B. Simon, Commun. Math. Phys. {\bf 27}, 1 (1972).
\bibitem {Ohkubo2002}
S. Ohkubo and K. Yamashita, Phys. Rev. C {\bf 66}, 021301(R) (2002).
\bibitem {Mahaux1986}% dispersion alpha+16O alpha+40Ca
C. Mahaux, H. Ngo, and G. R. Satchler, 
Nucl Phys. {\bf A449}, 354 (1986); 
C. Mahaux, H. Ngo, and G. R. Satchler, 
{\bf A456}, 134 (1986). 
\bibitem {Tohsaki1980} 
A. Tohsaki-Suzuki, M. Kamimura, and K. Ikeda,
Prog. Theor. Phys. Suupl. {\bf 68}, 359 (1980). 
\bibitem {Aoki1982} 
K. Aoki and H. Horiuchi, 
Prog. Theor. Phys. {\bf 67}, 1236 (1982).
\bibitem {Michel1989}
F. Michel, Y. Kondo, and G. Reidemeiter,
Phys. Lett. {\bf B 220}, 479 (1989).
\bibitem{Mohr2017}
P. Mohr, % 46,54Cr double folding model
Eur. Phys. J. A {\bf 53}, 209 (2017).
\bibitem {Abriola2008}%96Mo BE2 data
D. Abriola and A. A. Sozogni,
Nucl. Data Sheet {\bf 109}, 2501 (2008).
\bibitem{Mohr2007} % 104Te folding model
P. Mohr,
Eur. Phys. J. A {\bf 31}, 23 (2007).
\bibitem{Kiss2009} 
G. G. Kiss {\it et al.},
% P. Mohr, Zs. Ful?? op, D. Galaviz, Gy. Gyurky, ?? Z. Elekes, E. Somorjai, A. Kretschmer, K. Sonnabend, A. Zilges, and M. Avrigeanu
Phys. Rev. C {\bf 80}, 045807 (2009).
\bibitem{Abele1993}%20Ne, 19F folding 
H. Abele and G. Staudt, Phys. Rev. C {\bf 47}, 742 (1993).
\bibitem{Atzrott1996}%44Ti folding 
U. Atzrott, P. Mohr, H. Abele, C. Hillenmayer, and G. Staudt,
Phys. Rev. {\bf C} 53, 1336 (1996).
\bibitem{Hoyler1994} %212Po folding
F. Hoyler, P. Mohr, and G. Staudt, 
Phys. Rev. C {\bf 50}, 2631 (1994).
\bibitem{Saito1968}
S.~Saito, Prog.~Theor.~Phys. {\bf 40}, 893 (1968); 
S.~Saito, Prog.~Theor.~Phys. {\bf 41}, 705 (1969). 
\end{thebibliography}
\end{document}